\documentclass[twocolumn,showpacs,preprintnumbers,aps,pra]{revtex4}

\usepackage{graphicx}

\begin{document}

\title{Efficient and spectrally bright source of polarization-entangled photons}

\author{Friedrich K\"{o}nig,\footnote{Now at School of Physics and Astronomy,
University of St.\ Andrews, North Haugh, St. Andrews, Fife, KY16
9SS, Scotland.} Elliott J. Mason, Franco N. C. Wong, and Marius A. Albota}
\affiliation{Research Laboratory of Electronics, Massachusetts
Institute of Technology, Cambridge, Massachusetts 02139}

\date{\today}

\begin{abstract}
We demonstrate an efficient fiber-coupled source of nondegenerate polarization-entangled photons at 795 and 1609\,nm using bidirectionally pumped parametric down-conversion in bulk periodically poled lithium niobate.  The single-mode source has an inferred bandwidth of 50\,GHz and a spectral brightness of 300 pairs/s/GHz/mW of pump power that is suitable for narrowband applications such as entanglement transfer from photonic to atomic qubits.
\end{abstract}


\maketitle

\section{Introduction}
The demonstration of the Einstein-Podolsky-Rosen (EPR) paradox is one of the most striking quantum effects observed to date.  It reveals how entanglement of a nonseparable state produces nonlocal correlations that cannot be explained classically, as manifested in the violation of the Bell's inequalities \cite{CHSH}. Nonlocality is at the heart of a number of applications such as quantum cryptography and long-distance teleportation, which require efficient distribution of entanglement over long distances.  Photonic qubits are more easily transported over long distances than atomic qubits and, as a result, distribution of entangled  photons is an essential part of a support infrastructure for quantum communications and distributed quantum computation networks.

In this work we report on a source of polarization-entangled photon pairs that is suitable for long-distance quantum information processing.
Entangled photons at 795 and 1609\,nm are created by nondegenerate parametric
down-conversion with a spectral brightness that is sufficient for loading and entangling narrowband Rb-based quantum memories at 795\,nm.  The shorter-wavelength photon of this source can be used to directly excite a trapped Rb atom in a high-finesse cavity that serves as a local quantum memory. The other photon at 1609\,nm is suited for fiber-optic delivery because it lies in the low-loss transmission window of optical fibers (1500--1650\,nm).

In one proposed architecture for long-distance teleportation \cite{architecture}, a local and a remote Rb quantum memory are loaded using entangled photons. The stored atomic entanglement is then used to teleport an unknown atomic state from one location to the other \cite{Rbpaper}. Implicitly required for the architecture are the efficient transport of entangled photons over long distances and wavelength matching between the entangled photons and the trapped atomic Rb. Because of unavoidable losses through long optical fibers and the narrow bandwidth of atomic Rb of tens of MHz, it is essential to have a spectrally bright source of entangled photons at the above mentioned wavelengths.  Note, however, that the entanglement of successfully delivered qubits is not degraded in this scheme \cite{architecture}. For loading of the remote quantum memory, the 1609-nm photon has to be upconverted to the Rb wavelength at 795\,nm with preservation of its polarization quantum state. 90\%-efficient upconversion \cite{upconversion} has recently been demonstrated at the single-photon level for a fixed input polarization, and quantum-state preserving upconversion (for an arbitrary polarization qubit) is expected to have a similar conversion efficiency.

Typical down-conversion sources using beta barium borate (BBO) crystals
\cite{Kwiat95,Kwiat99,Weinfurter01} are not suitable for narrowband applications such as quantum memory loading because of their large bandwidths of 1--10 nm. These sources also have relatively low flux, partly because of inefficient collection of the output photons in their emission cone and partly because of the small nonlinear coefficient of BBO\@.  Down-conversion in waveguide nonlinear crystals \cite{Gisin01,Walmsley01} are much more efficient but the outputs are still broadband and polarization entanglement using these waveguide crystals has not been reported.

We have previously demonstrated a high-flux source of nondegenerate photon pairs at 795 and 1609\,nm in bulk periodically poled lithium niobate (PPLN) in a collinearly propagating geometry \cite{PPLNdownconversion}.  In this work we have improved the source brightness to enter a regime suitable for narrowband applicatons such as atomic excitation and we have modified the apparatus for the generation of polarization-entangled photon pairs.  PPLN has an effective nonlinear coefficient that is an order of magnitude larger than that of BBO and it can be tailored to phase match at any set of operating wavelengths within the transparency window of the crystal.  In addition, PPLN has a wide temperature tuning range and is commercially available.  Compared with more conventional non-collinear configurations \cite{Kwiat95,Kwiat99,Weinfurter01}, collinear propagation in noncritically angle phase-matched down-conversion permits the use of a long crystal for more efficient generation.  Moreover, the beam-like (instead of cone-like) output can be collected and fiber-coupled more efficiently for long-distance transport.  This collinear geometry has recently been utilized in a type-II phase-matched periodically poled KTiOPO$_4$ (PPKTP) down-converter to efficiently generate polarization-entangled photons at 795\,nm \cite{PPKTP1}.  In a dual-pumping scheme with interferometric combination of the collinear type-II phase-matched PPKTP outputs a tenfold increase in flux has been demonstrated \cite{PPKTP2}.  This dual-pumping method employs a single nonlinear crystal that is driven coherently by two counter-propagating pumps and is suitable for nondegenerate operation with wavelength tuning.

For the current work, we have applied the dual-pumping technique to the generation of highly nondegenerate polarization-entangled photons using PPLN as the nonlinear crystal.  By coupling the output signal and idler photons into their respective single-mode optical fibers, we obtain single-spatial-mode photon pairs that are highly suited for spatial mode matching and fiber-optic transport of the 1.6-$\mu$m light. Single-mode fiber-optic delivery is particularly useful for applications such as the above-mentioned 
quantum-memory loading of a trapped atom in a high-finesse optical cavity.  Fiber coupling of the outputs also allows one to select a mode-matched portion of the spontaneously emitted output field from the crystal in order to limit the spectral bandwidth of the light. In our case, we have obtained a spectral bandwidth of $\sim$50\,GHz which is less than the phase-matching bandwidth ($\sim$150\,GHz) of our 2-cm-long PPLN crystal.  We have achieved a spectral brightness that is suitable for narrowband wavelength-sensitive applications such as the proposed long-distance teleportation protocol \cite{architecture}.

In the next section we describe our dual-pumping generation scheme for the creation of polarization-entangled photon pairs in PPLN\@. In Section 3, the characteristics of the PPLN source are presented when operated as a single-pass down-converter. Finally, in Section 4 we describe the bidirectional pumping scheme and the violation of Bell's inequality as a measure of the purity of the polarization entanglement, before we conclude in Section 5.

\section{Generation of polarization entanglement}
Spontaneous parametric down-conversion (SPDC) in a nonlinear optical crystal is a standard technique for generating polarization-entangled photons.  In SPDC, a pump photon is converted into two subharmonic photons, called signal ($S$) and idler ($I$), that have definite polarizations.  For example, in a low-gain type-I phase-matched interaction, the signal and idler photons are co-polarized with an output state (after ignoring the vacuum and higher-order modes) given by $|\Psi\rangle =| 1 \rangle_{S\bullet} | 1
\rangle_{I\bullet}$, where $\bullet$ refers to horizontal polarization.  In this example, $|\Psi\rangle$ is a separable state and there is no polarization entanglement.  It is clear that SPDC, by itself and without additional arrangements, does not produce polarization entanglement in which each photon of the down-converted pair appears to be randomly polarized and yet the pair is polarization correlated.

\begin{figure}
\centerline{\scalebox{0.45}{\includegraphics{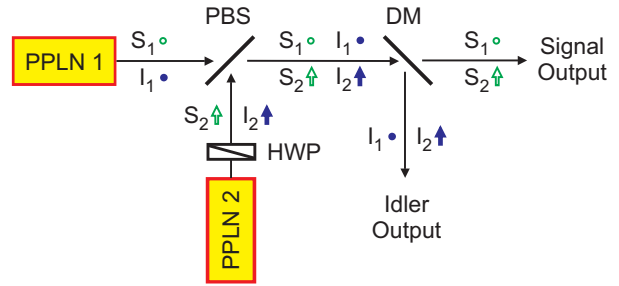}}}
\caption{Schematic view of two coherently driven and combined PPLN down-converters. Horizontal (vertical) polarization is denoted by $\bullet$ ($\uparrow$). HWP, half-wave plate; PBS, polarizing beam splitter; DM, dichroic mirror.} \label{stategeneration}
\end{figure}

It was proposed \cite{Kwiat94,Shapiro00} and recently demonstrated \cite{PPKTP2} in a type-II phase-matched system that polarization entanglement can be readily obtained if the outputs from two coherently driven identical down-converters are interferometrically combined.  A remarkable property of such a bidirectionally pumped arrangement is that spatial, spectral, or temporal filtering is no longer necessary for obtaining high quality entanglement \cite{PPKTP2}. In our current experiment, we explore a type-I phase-matched system in a similar bidirectional pumping arrangement. Since PPLN is type-I phase-matched the signal and idler outputs have the same polarization, and hence they can only be separated if the outputs are very different in wavelength, using a dispersing prism or dichroic mirror, for example.  A configuration that is suitable for nondegenerate type-I phase-matched down-conversion in PPLN is schematically shown in Fig.~\ref{stategeneration}. The horizontally polarized signal and idler fields in one direction (path 1) are combined with the vertically polarized signal and idler fields (after a $\pi/2$ polarization rotation) in the other direction (path 2) at the polarizing beam combiner.  The signal and idler outputs are then separated at the dichroic mirror.

After the interferometric combination, the lowest-order non-vacuum output is the biphoton state
\begin{eqnarray}
|\Psi\rangle &=&  \alpha | 1 \rangle_{S\bullet} | 1 \rangle_{I\bullet} | 0
\rangle_{S\uparrow} | 0 \rangle_{I\uparrow} + \beta | 0 \rangle_{S\bullet} | 0
\rangle_{I\bullet} | 1 \rangle_{S\uparrow} | 1
 \rangle_{I\uparrow} \nonumber \\
 &=& \alpha | H \rangle_S | H \rangle_I + \beta | V \rangle_S | V \rangle_I\,.
\label{state}
\end{eqnarray}
The complex coefficients $\alpha$ and $\beta$ represent the field strengths for the two down-converters and $\uparrow$ indicates vertical polarization. For a fiber-coupled implementation of Fig.~\ref{stategeneration}, $|\alpha|^2$ ($|\beta|^2 = 1-|\alpha|^2$) is proportional to the overall efficiency for pair generation in path 1 (2) direction, and subsequent propagation and coupling into the signal and idler fibers.  The phase of $\alpha$ ($\beta$) is the sum of the phases of the horizontally (vertically) polarized
signal and idler fields along their paths from the crystal to the dichroic mirror.  Note that the common path for the signal and idler fields after the polarizing beam combiner adds a common phase to $\alpha$ and $\beta$, which yields an inconsequential overall phase for the state $|\Psi \rangle$.  In Eq.~(\ref{state}), we have simplified the notation by not displaying the vacuum modes and by replacing $| 1 \rangle_\bullet$ and $| 1 \rangle_\uparrow$ with $| H \rangle$ and $| V \rangle$, respectively. Implicit in Eq.~(\ref{state}) is that the signal and idler photons have different wavelengths that allow them to be easily separated and individually manipulated. A key characteristic of this dual-pump configuration is that, with a proper choice of polarization basis, phases, and relative pumping strengths, one can in principle obtain any one of the four Bell states, independent of propagation or coupling losses. For instance, with $\alpha= - \beta = 1/\sqrt{2}$ and a $90^\circ$ rotation of the idler state, the singlet state $| \Psi^- \rangle = \alpha [| H \rangle_S | V \rangle_I -  | V \rangle_S | H \rangle_I ]$ is obtained.

\section{Single-pass down-conversion}
Figure 2 shows a schematic of our experimental setup for implementing the polarization entanglement source with two identical down-converters.  By using two counter-propagating pump beams to coherently drive a single PPLN crystal, the two down-converters can be made to be nearly identical (limited by the crystal grating uniformity).  We first describe and characterize our PPLN down-converter in single-pass operation with only one pump before detailing the coherent combining technique in the next section.

\begin{figure}
\centerline{\scalebox{0.42}{\includegraphics{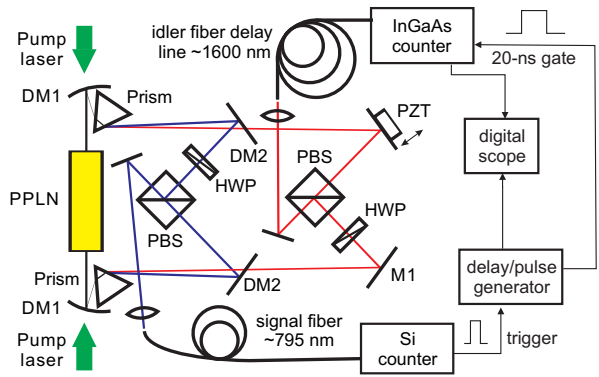}}}
\caption{Schematic of experimental setup used for single-pass and dual-pumped
down-conversion. DM, dichroic mirror; HWP, half-wave plate; PBS, polarizing beam
splitter; PZT, piezoelectric transducer.} \label{setup}
\end{figure}

We employed a 2-cm long and 0.5-mm thick PPLN crystal with a grating period of
21.6\,$\mu$m as the nonlinear medium.  We previously fabricated the crystal for type-I third-order quasi-phase-matched SPDC to generate tunable signal and idler photons centered at $\sim$0.8 and $\sim$1.6\,$\mu$m, respectively, from a continuous-wave (cw) pump at 532\,nm \cite{PPLNdownconversion}.  The crystal was antireflection coated at the pump, signal, and idler wavelengths with reflectivities of $<$1\%, $<$1\%, $\sim$8\% per surface, respectively.  Using difference-frequency generation (DFG) we observed a phase-matching bandwidth of $\sim$150\,GHz at a fixed pump wavelength and a fixed crystal temperature.  By varying the crystal temperature, we were able to tune the outputs over tens of nm, with a tuning coefficient of $\sim$150\,GHz/$^\circ$C\@.   For outputs at 795 and 1609\,nm, we operated the PPLN crystal at $176 \pm 0.1^\circ$C\@.  From the DFG results we estimate the effective third-order nonlinear coefficient to be 3.8\,pm/V, which is less than expected due to grating imperfections and mode mismatch in the DFG measurements.

The PPLN down-conversion source, including the fiber-coupled pump input and the output fiber couplers, was set up on a $38 \times 74$\,cm$^2$ breadboard for improved mechanical stability and portability. The 532-nm pump was derived from a cw frequency-doubled Nd:YVO$_4$ laser (Coherent Verdi-8) with a maximum output power of 8\,W\@.  We coupled a small fraction of the laser power into a single-mode polarization maintaining fiber which delivered the pump light to the PPLN source. This pump fiber exhibited slight multimode behavior and the output was mode cleaned with a pinhole and then focused into the center of the crystal with a beam waist of 90\,$\mu$m. Typical pump powers incident upon the crystal were 1\,mW; however, the crystal could easily be pumped by a few hundred mW of cw power without damage.

The dichroic curved mirrors DM1 (highly reflecting at 0.8 and 1.6\,$\mu$m and highly transmitting at 532 nm) in Fig.~\ref{setup} served as focusing lenses for the pump and to remove the pump from the down-converted output at the exit.  The down-converted light was spectrally separated into signal and idler beams with a
dispersing prism which also steered the residual pump light away from the signal and idler paths. Signal and idler beams could then be individually manipulated before being coupled into their respective single-mode optical fibers for transport or photon counting detection.

\begin{figure}
\centerline{\scalebox{0.25}{\includegraphics{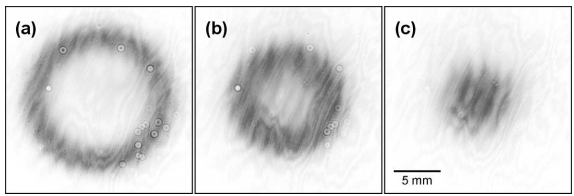}}} 
\caption{Far-field CCD camera images of spontaneously down-converted light through a 1-nm interference filter centered at 795\,nm. PPLN temperatures were (a) $177.6^\circ$C, (b) $180.6^\circ$C, and (c) $183.6^\circ$C. The 795-nm light was phase-matched at different angles as a function of temperature. Fringing is due to camera etalon effects.} \label{figring}
\end{figure}

First we investigated the spontaneous emission characteristics of the bulk PPLN crystal.  We passed the signal output through a 1-nm interference filter (IF) centered at 795\,nm and imaged the output onto a high-resolution charged-coupled device (CCD) camera with a detection sensitivity of a few photons. Figure~\ref{figring} shows the evolution of the 795-nm output as a function of the crystal temperature from a cone-like pattern, as in noncollinear SPDC, to a beam-like output, as in a collinear geometry.  Figure~\ref{figring}(a) shows a typical ring pattern of the observed signal output emerging from the crystal.  The ring diameter is directly related to the internal emission angle at which the output at the 795-nm wavelength was phase-matched.  By varying the crystal temperature the phase-matching angle for 795-nm emission was modified as shown in Fig.~\ref{figring}(a-c). In Fig.~\ref{figring}(c) the output was collinear with a phase matching temperature $T = 183.6^\circ$C.

\begin{figure}
\centerline{\scalebox{0.35}{\includegraphics{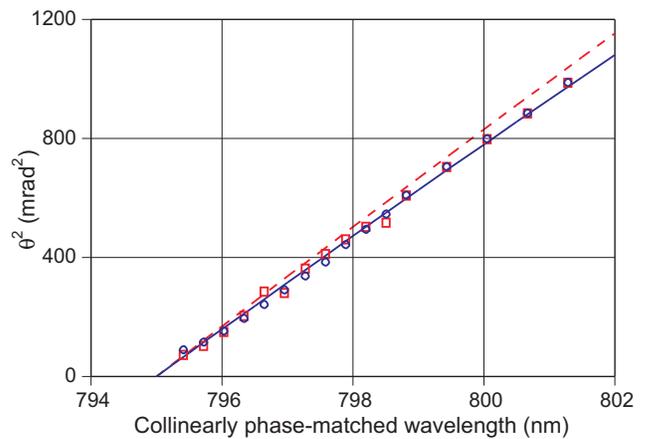}}}
\caption{Plot of external far-field PPLN emission angles, as measured by the ring diameters of CCD images along the crystal's $z$-axis (open circles) and $y$-axis (open squares) as a function of the collinearly phase-matched wavelength $\lambda_{CPM}(T)$, where $T$ is the PPLN crystal temperature. The solid and dashed lines are the theoretical emission angle curves corresponding to the crystal's $y$ and $z$ axes, respectively, obtained from the PPLN's Sellmeier equations with no adjustable parameters.}
\label{theory}
\end{figure}

We measured the ring diameters and hence the phase-matching angles of the 795-nm PPLN emission as a function of the crystal temperature.  At each temperature $T$ we obtained the corresponding collinearly phase-matched signal wavelength $\lambda_{CPM}(T)$ by DFG phase-matching measurements. Figure~\ref{theory} shows the measured emission angles $\theta(T)$ relative to $\lambda_{CPM}(T)$, and compares them with the computed phase-matching angles obtained from the Sellmeier equations \cite{Byer91} for noncollinearly phase-matched PPLN\@. The excellent agreement between the experimentally observed signal emission angles and the theoretical values that we obtain without free
parameters allows us to infer the mode characteristics of the idler output beam.  In particular, because of the signal-idler wavelength difference, the 1.6-$\mu$m idler emission angle is twice the 0.8-$\mu$m signal angle, and the idler bandwidth in nm is four times as large as the corresponding signal bandwidth of the signal-idler pair emission.

\begin{figure}
\centerline{\scalebox{0.35}{\includegraphics{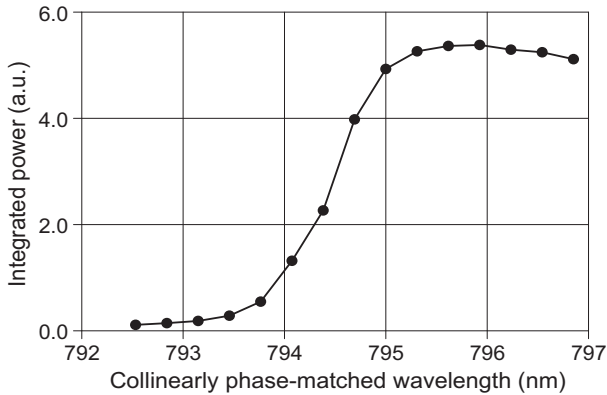}}}
\caption{Integrated power of output emission at 795\,nm in a 1-nm bandwidth.  PPLN crystal temperature $T$ was set for collinear phase matching at the wavelength $\lambda_{CPM}(T)$. The total output power shows little dependence on the emission angles for $\lambda_{CPM} \geq 795$\,nm.} \label{ringpower}
\end{figure}

The CCD camera was used to measure the emitted signal power within the 1-nm IF bandwidth.  We integrated the pixel output voltages spatially over the ring structure and subtracted the background (obtained without input light) from it. The 795-nm signal power was measured as a function of crystal temperature within a 1-nm bandwidth.  In Fig.~\ref{ringpower} we plot the power versus the wavelength $\lambda_{CPM}(T)$, at which collinear phase matching was achieved at temperature $T$.  Below $\lambda_{CPM} \approx 794$\,nm, the 795-nm output was not phase-matched and the emission vanished.  Beyond $\lambda_{CPM}=795$\,nm, the output power remains more or less constant, indicating that the integrated power was only weakly dependent on the phase-matching temperatures (and emission angles).

Next we optimized the coupling efficiency of the photon pairs into single-mode fibers by utilizing the mode structure information. In particular, we found that PPLN temperature tuning was  a useful tool to vary slightly the down-converted beam parameters without affecting the down-conversion efficiency.  We measured the fiber coupling efficiencies at three different 795-nm IF bandwidths of 0.11, 1, and 3\,nm for the signal photons. The PPLN temperature was nominally set for collinear phase matching at $\lambda_{CPM} = 795$\,nm  and aspheric lenses of various focal lengths were used at each IF bandwidth setting.  The signal photon flux before the signal fiber was measured with a high-sensitivity Si photodiode under strong pumping.  For measuring the flux after the single-mode fiber we reduced the pump power and used a Si single-photon counting module (SPCM). After accounting for various losses and adjusting for the different pumping levels and different detection efficiencies for the photodiode and the single-photon counter, the best fiber coupling efficiency of 18\% for the signal beam was obtained using the 0.11-nm IF\@.  In contrast, with the 3\,nm IF the coupling was at least an order of magnitude less efficient due to the much larger angular bandwidth of the output signal photons.

For a 3-nm bandwidth, we infer from the measurements a pair generation rate of $2 \times 10^7$\,/s/mW of pump power at the output of the crystal. The single-mode fiber-coupled signal rate (corrected for detector efficiency) was $5.1 \times 10^4$\,/s/mW of pump power within the same 3-nm spectral bandwidth. This rate was reduced from that at the crystal because of propagation and fiber coupling losses.

We then studied the conditional detection probabilities by making signal-idler
coincidence measurements, as shown in the schematic setup of Fig.~\ref{setup}.  The 800-nm signal fiber was connected to a fiber-coupled Si SPCM (PerkinElmer AQR-13-FC) which produced a timing trigger upon detection of an incoming signal
photon.  A time-delay pulse generator then produced a bias gating pulse of 7--20 ns duration which was sent to a Peltier-cooled passively-quenched InGaAs avalanche photodiode (APD) operating in Geiger mode for the detection of the conjugate 1609-nm idler photon.  The design and properties of our home-made InGaAs APD single-photon counter have been described elsewhere \cite{PPLNdownconversion}. The main characteristics of this InGaAs single-photon counter are a dark count probability of 0.1$\%$ per 20-ns gate and a quantum efficiency of 19.8\% at a detector temperature of $-50^\circ$\,C\@.   The generated electrical pulses from the detectors were then recorded by a computer and their timing information was used to identify firings of the Si and InGaAs counters corresponding to pair coincidences.  Typically we used a coincidence window of 4\,ns.

The conditional probability of detecting an idler photon, given the detection of a signal photon, was 9.4\% using the narrowband 0.11-nm IF\@.   In calculating this conditional probability we subtracted the background in the coincidence rate due to accidental and dark counts (probability of $\sim$0.04\%).  This background rate was measured by offsetting the start of the gating pulse for the InGaAs counter by an amount greater than the 4-ns coincidence window. Given the 19.8\%-efficient InGaAs counter, this implies $\sim$50\% losses in idler propagation, mode matching, and fiber coupling into the 1.6-$\mu$m optical fiber, conditioned on signal photon detection. Single-mode pair generation is best characterized by the rate of photon pairs that are successfully coupled into their respective single-mode fibers, i.e., the detected rate corrected for detector quantum efficiencies. In the case of the 0.11-nm IF, we observed a signal-idler coincidence rate of 450\,/s/mW, from which we infer a single-mode fiber-coupled pair flux of $\sim$4100\,/s/mW before detection. Without an IF the conditional probability decreased to 5.2\% because additional spatial modes were coupling into the signal fiber whose idler counterparts could not be all coupled simultaneously into the idler fiber in an efficient way. However, without using spectral filters we note that there were many more signal photons coupled into the signal fiber (partly because of the absence of the propagation loss through the IF)  and we measured a single-mode fiber-coupled pair flux of 16\,000\,/s/mW before detection in this case.

\section{Polarization entanglement by bidirectional pumping}
Coherent combination of two down-converters for the generation of polarization
entanglement shown schematically in Fig.~\ref{stategeneration} was implemented with the setup of Fig.~\ref{setup} which we have described for the single-pass case in the previous section. To ensure that the two down-converters were as identical as possible, a single PPLN crystal was pumped in counter-propagating directions along the crystal's $x$-axis.  The fiber-coupled 532-nm pump light was split into two beams with a 50--50 beam splitter.  Each of the pump beams was directed into the PPLN crystal as shown in Fig.~\ref{setup} and the two outputs followed nearly identical paths as described in the previous section.  In order to generate polarization entanglement, the signal and idler fields of one of the down-converters were rotated by 90$^\circ$ with half-wave plates (HWPs) before combining with the output from the other down-converter.

The spectral contents of the two down-converters were identical by our use of a single crystal.  However, spatial and temporal mode matching were still necessary because there were four path lengths (two for the signals and two for the idlers) that must be properly adjusted for efficient generation of polarization entanglement. When the two signal fields were combined at the polarizing beam splitter (PBS), the two spatial modes should be identical to avoid spatial-mode distinguishability, and similarly for the idler fields. This was accomplished by making sure that the two pump foci inside the crystal were the same in size and location, and the two signal and two idler path lengths were the same.  The tolerance of the path length difference (a few mm) was dictated by the confocal parameters at the location of the PBS\@. Note that this mode overlap facilitated equal coupling to the respective signal and idler fibers, where the fields were projected into a single spatial mode. A more stringent requirement is temporal indistinguishability. That is, the path length difference between the two combining signal fields should be the same as that between the two idler fields within the coherence length of the photons, which was determined by the bandwidths of the phase matching, the IF, and the fiber coupling.

The photon bandwidth can be easily determined in a Hong-Ou-Mandel (HOM) interferometric measurement for degenerate signal and idler wavelengths. For nondegenerate wavelengths two-photon quantum interference visibility can be similarly utilized to measure the photon bandwidth.  Quantum interference is also the basis for demonstrating Bell's inequality in the Clauser-Horne-Shimony-Holt (CHSH) form \cite{CHSH} that yields a measure of
the quality of polarization entanglement.

The biphoton output state of the bidirectionally pumped down-converter, for equal pump power in both directions, is
\begin{equation}
|\Psi\rangle = (| H \rangle_S | V \rangle_I - e^{i\phi} | V
\rangle_S | H \rangle_I)/\sqrt{2}\,, \label{outputstate}
\end{equation}
where $\phi$ is the relative phase that is a function of the phase difference between the two pumps (hence coherent pumping is required), and the signal and idler phases acquired along the four path lengths.  In the experiment, we controlled $\phi$ by a mirror mounted on a piezoelectric transducer (PZT) in one of the idler paths, as shown in Fig.~\ref{setup}.  We note that the output state is a singlet polarization-entangled state for $\phi = 0$.

For quantum interference observation and analysis, we installed polarization analyzers (a HWP followed by a horizontally-transmitting PBS) at the entrances to the signal and idler single-mode optical fibers.  The two analyzers were initially set parallel and oriented to transmit an equal amount of light from the two pumping directions (equivalent to transmission at 45$^\circ$ relative to the horizontal polarization).  All angles of our analyzers were measured with respect to this initial orientation.

\begin{figure}
\centerline{\scalebox{0.33}{\includegraphics{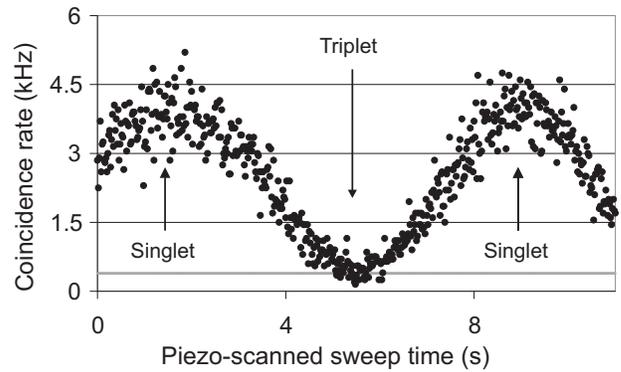}}}
\caption{Two-photon quantum interference from the bidirectionally pumped down-converter, showing the detected coincidence rate as a function of the PZT mirror sweep with crossed analyzers.  The PZT changes the relative phase $\phi$ of the output state of Eq.~(\ref{outputstate}), yielding a maximum rate for $\phi = 0$ (singlet) and a minimum rate for $\phi = \pi$ (triplet).  See text for experimental details.} \label{quantuminterference}
\end{figure}

We first demonstrated the generation of different two-photon states by varying the PZT mirror position, as manifested in the two-photon quantum interference.  With the idler analyzer set orthogonal to the signal one, the PZT was swept and the two-photon output state went through the singlet state at $\phi=0$ and the triplet state at $\phi=\pi$.  We observed a sinusoidal modulation in the coincidence detection rate as a function of the PZT sweep, as shown in Fig.~\ref{quantuminterference}.  At $\phi=0$, the singlet-state output is obtained which is invariant under coordinate rotation and signal and idler are always orthogonally polarized. Hence, a maximum in the coincidence rate was obtained for $\phi=0$. At $\phi=\pi$, the output state was transformed by the HWP into a triplet state with parallel polarizations for signal and idler, thus yielding a minimum in the coincidence rate, as displayed in Fig.~\ref{quantuminterference}. For non-entangled photon pairs, the sweep would not produce any modulation in the coincidence rate, which would lie mid-way between the maximum and minimum values.  We obtain a quantum-interference visibility of $\sim$93\% if dark counts and accidentals are
subtracted (grey line in Fig.~\ref{quantuminterference}). The quantum interference coincidence measurements in Fig.~\ref{quantuminterference} were made with $\sim$80\,mW of total pump power and the data were taken with the 0.11-nm IF. The interferometric setup was very stable such that no locking was required. The observed free-running phase drift after 5 minutes was much lees than $\pi/4$.  We used a high-speed multichannel analyzer (PicoQuant's TimeHarp 200) for time-resolved coincidence counting with an effective coincidence window of 4\,ns.  As the PZT was slowly swept the computerized data collection recorded the number of coincidences in 20-ms measurement time intervals. Note that the noise in Fig.~\ref{quantuminterference} is Poissonian if we consider that the data were sampled in 20-ms time intervals. This initial set of measurements permitted us to further optimize the quantum-interference visibility and the coincidence rate.

\begin{figure}
\centerline{\scalebox{0.33}{\includegraphics{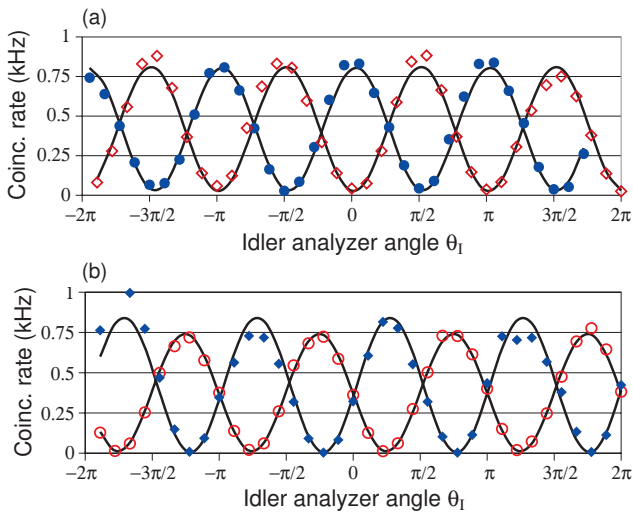}}}
\caption{Plot of coincidence rate as a function of idler polarization analyzer angle $\theta_I$, where  at $\theta_I = 0$, the analyzer was set to couple equal amounts of idler from the two down-converters.  (a) Signal polarization analyzer was set at $\theta_S = 0$ (passing equal amounts of signal light from the two
down-converters) for the singlet output state $\phi = 0$ (open diamond) and the triplet output state $\phi = \pi$ (solid circle). (b) $\theta_S = \pm\pi/4$ to pass the output from the single down-converter pumped from below (open circle) or above (solid diamond).  Below and above refers to pumping directions as
indicated in Fig.~\ref{setup}.} \label{moredata}
\end{figure}

Additional quantum interference measurements with the 0.11-nm IF were carried out and are shown in Fig.~\ref{moredata}. In Fig.~\ref{moredata}(a) the coincidence counts are plotted for the singlet (open diamond) and triplet (solid circle) states as a function of the idler analyzer angle $\theta_I$ when the signal analyzer angle $\theta_S$ was set to equally transmit light from both of the down-converters ($\theta_S = 0$). Here we obtain quantum-interference visibility of 94\% for the singlet and triplet in Fig.~\ref{moredata}(a).  In Fig.~\ref{moredata}(b) a similar plot of the coincidence counts is made with the signal analyzer transmitting signal photons from the lower (open circle) and upper (solid diamond) pumping beams (Fig.~\ref{setup}), with $\theta_S =\pi/4$ and $-\pi/4$, respectively. By setting the signal analyzer $\theta_S$ to pass the signal photons of just one of the down-converters, we were simply measuring the pair generation rate as a function of $\theta_I$. For the data in Fig.~\ref{moredata}(b), we obtained a higher visibility of 99.8\% for the horizontal ($\theta_S =-\pi/4$) and 98\% for the vertical ($\theta_S = \pi/4$) orientation of the signal analyzer.  We believe that the visibilities in Fig.~\ref{moredata}(a) were not limited by spatial mode mismatch, temporal mismatch, or a rate imbalance of the two down-converters because the interfering signal fields shared the single mode of the fiber and temporal overlap and rate balance were adjusted accurately. Therefore we conclude that a spectral distinguishability between the two down-converters is the most likely reason, caused by either a difference in the spectra of the fiber-coupled signals and idlers or the crystal nonuniformity along its length.

We have used the quantum interference signature to measure the photon bandwidth of the fiber-coupled photon pair in the absence of an IF\@.  By translating the mirror M1 over a distance of a few mm in fine steps, we monitored the interference visibility as a function of distance. If the two orthogonal idler fields do not overlap in time, relative to the time difference of the two orthogonal signal fields, then the two-photon quantum interference disappears. We observed that the full distance at half of the peak visibility was 3.4\,mm, or $\sim$11\,ps in time.  If we assume that the fiber coupling acts as a Gaussian filter, then the corresponding fiber-coupled photon pulse width is 8.8\,ps, or 50\,GHz in spectral bandwidth. This narrow bandwidth was
obtained for a particular mode matching configuration which efficiently coupled only the light of this small bandwidth. This is a manifestation of the high degree of spatial and spectral correlation in the down-converted output. We have also performed a series of quantum interference measurements to demonstrate the violation of Bell's inequality in the CHSH form \cite{CHSH}. We follow the standard procedure of measuring coincidence rates at different polarization analyzer angles using the singlet state as the input. The drift of the dual-pumped down-converter interferometer was found to be negligible as it did not degrade the interference over the 6-minute measurement time, which included manually setting the analyzer angles. We measured an $S$ parameter of $2.606\pm 0.010$ for a pump power of 2.2\,mW per beam.  The results indicate a violation of 60 standard deviations over the classical limit of 2. Perfect polarization entanglement would have yielded maximum violation of the Bell's inequality with $S=2\sqrt{2}$.

\section{Summary}
We have demonstrated an efficient source of polarization-entangled photons at highly nondegenerate wavelengths using bidirectional pumping of a single PPLN crystal followed by coherent combination of the down-converted outputs.  Our source is fiber-coupled, widely tunable by tens of nm in wavelength via temperature tuning of the PPLN crystal, and takes advantage of a convenient, collinearly propagating geometry.  Using third-order, type-I quasi-phase matching in a 2-cm long PPLN crystal, we have obtained a fiber-coupled polarization-entangled pair flux of $\sim$16\,000\,/s/mW of 532-nm pump power in a bandwidth of 50 GHz.  For a bandwidth of 50 MHz, our source has a flux of $\sim$16\,/s/mW.  This number can be scaled up by three orders of magnitude using a first order grating (factor of 9 improvement) in PPLN with a better quality grating duty cycle (factor of 20\%), and a higher pump power of 100 mW or more.  These simple improvements can be implemented without changing the experimental configuration and would yield a photonic source of polarization entanglement suitable for entanglement transfer to atoms via direct atomic excitations which have bandwidths of tens of MHz. The source could be used for testing long-distance teleportation schemes, in which the entanglement is stored in trapped-atom quantum memories \cite{architecture}.  The bidirectional pumping configuration in our bulk PPLN system can be applied to a PPLN waveguide down-converter, thereby combining the significant efficiency improvement of a waveguide configuration with the ease of generating polarization entanglement.  Furthermore, the collinear geometry is suitable for implementing a cavity-enhanced parametric amplifier configuration \cite{Shapiro00}, which should yield polarization-entangled photons that have bandwidths closely matching those of atomic excitations.  Our source therefore is useful for a number of quantum information processing tasks such as long-distance teleportation and quantum cryptography. 

\section{Acknowledgment}
This work was supported by the DoD Multidisciplinary University Research Initiative (MURI) program administered by the Office of Naval Research under Grant N00014-02-1-0717.  We acknowledge substantial laboratory help from Christopher Kuklewicz.

\end{document}